\documentclass[reprint,superscriptaddress,preprintnumbers,nofootinbib, amsmath,amssymb, prdfloatfix, prl]{revtex4-1}
\usepackage{CJK}
\usepackage{graphicx}
\usepackage{dcolumn}
\usepackage{upgreek}
\usepackage{bm}
\usepackage[colorlinks=true,pdfstartview=FitV,breaklinks=true]{hyperref}
\usepackage[dvipsnames,table]{xcolor}
\hypersetup{urlcolor=BlueViolet,
	    citecolor=Plum,
	    linkcolor=PineGreen}
\usepackage{lipsum}	    
\usepackage{titlesec}
\titlespacing*{\section}{0pt}{0.8\baselineskip}{0.8\baselineskip}
\usepackage{times}

\begin{document}

\setlength{\abovedisplayskip}{4pt}
\setlength{\belowdisplayskip}{4pt}

\preprint{CPPC-2021-04}

\title{Phenomenology of the companion-axion model: photon couplings}

\author{Zhe Chen}
\email{zche8090@uni.sydney.edu.au}
\affiliation{Sydney Consortium for Particle Physics and Cosmology, School of Physics, The University of Sydney, NSW 2006, Australia }
\author{Archil Kobakhidze}
 \email{archil.kobakhidze@sydney.edu.au}
\affiliation{Sydney Consortium for Particle Physics and Cosmology, School of Physics, The University of Sydney, NSW 2006, Australia }

 \author{Ciaran A. J. O'Hare}
 \email{ciaran.ohare@sydney.edu.au}
\affiliation{School of Physics, The University of Sydney and ARC Centre of Excellence for Dark Matter Particle Physics, NSW 2006, Australia }

 \author{Zachary S. C. Picker}
 \email{zachary.picker@sydney.edu.au}
\affiliation{School of Physics, The University of Sydney and ARC Centre of Excellence for Dark Matter Particle Physics, NSW 2006, Australia }

 \author{Giovanni Pierobon}
 \email{g.pierobon@unsw.edu.au}
 \affiliation{School of Physics, The University of New South Wales, Sydney NSW 2052, Australia}

\begin{abstract}
We study the phenomenology of the `companion-axion model' consisting of two coupled QCD axions. The second axion is required to rescue the Peccei-Quinn solution to the strong-CP problem from the effects of colored gravitational instantons. We investigate here the combined phenomenology of axion-axion and axion-photon interactions, recasting present and future single-axion bounds onto the companion-axion parameter space. Most remarkably, we predict that future axion searches with haloscopes and helioscopes may well discover two QCD axions, perhaps even within the same experiment.
\end{abstract}

\maketitle

\textbf{\textit{Introduction}}.---By far the most enduring solution to the strong-CP problem of quantum chromodynamics (QCD) is the theory first proposed by Peccei and Quinn (PQ)~\cite{Peccei:1977hh, Peccei:1977ur}. In their model, the dynamical degree of freedom provided by a spontaneously broken global $U(1)_{\rm PQ}$ symmetry is used to cancel the unobserved~\cite{Abel:2020gbr} CP-violating term in the QCD Lagrangian. The theory is simple and elegant, and the predicted pseudo Nambu-Goldstone boson---the `axion'---can simultaneously constitute the dark matter (DM) that pervades our Universe~\cite{Preskill:1982cy,Abbott:1982af,Dine:1982ah,Marsh:2015xka}. 

However, the axion solution to the strong CP problem can be spoiled in the presence of additional sources of PQ symmetry breaking. As it turns out, colored gravitational instantons are just such a source~\cite{Chen:2021jcb}. These are a generic prediction of the Standard Model with General Relativity so their contribution cannot be ignored---but the situation would be the same if instantons of some other confining gauge theory were present as well.

\begin{figure}[t]
\begin{center}
\includegraphics[trim = 0mm 0mm 0mm 0mm, clip, width=0.49\textwidth]{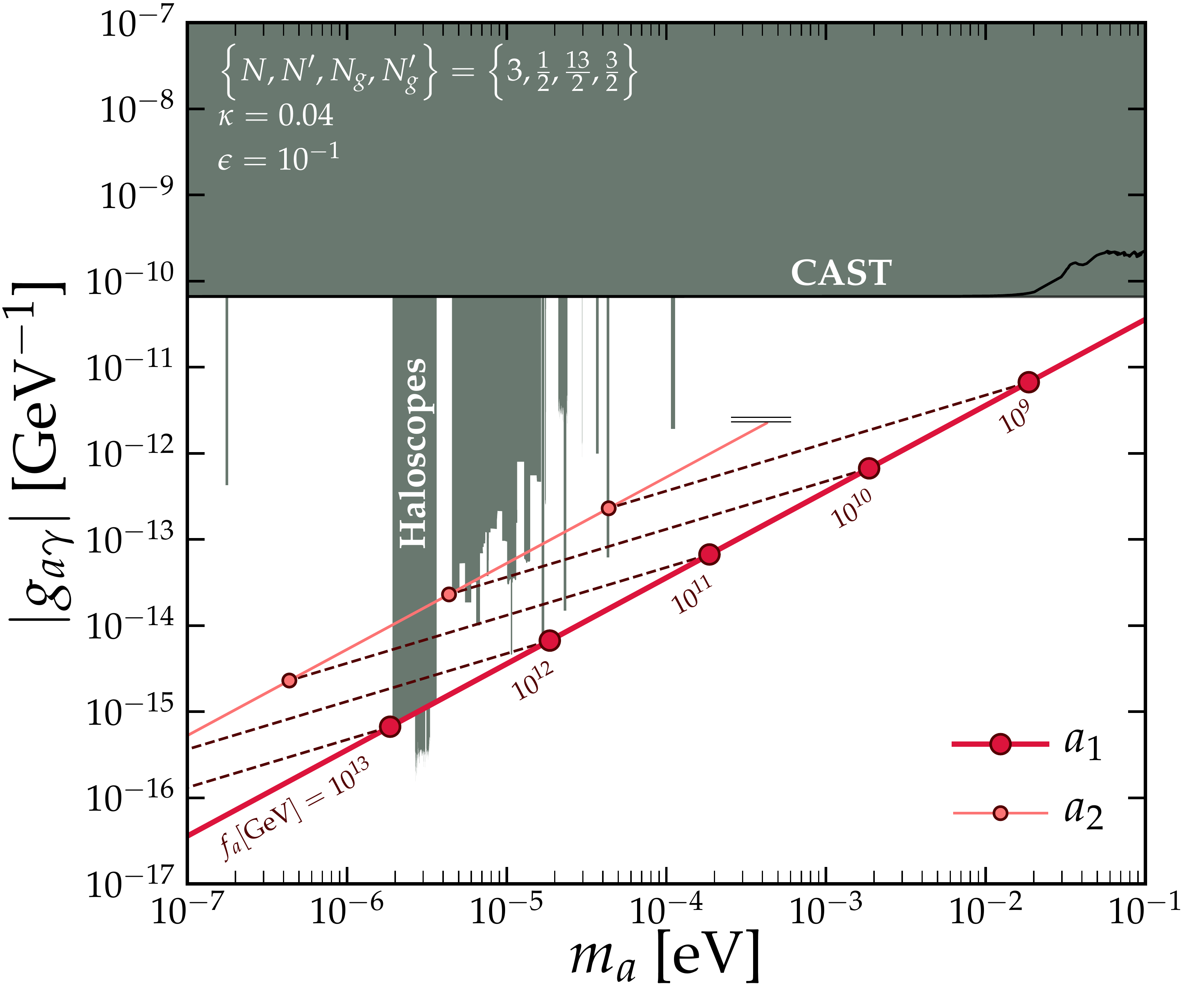}
\caption{Photon coupling-mass relation for the two axion mass eigenstates $a_{1,2}$. We draw a dashed line to emphasize the connection between the two axions which is set by the value of the free parameter $\epsilon = f_a/f^\prime_a$. The lighter axion's line cuts off at around $m_2\sim$~0.4~meV when its heavier companion is excluded by stellar cooling. In gray we display some existing constraints from haloscope searches for DM~\cite{Asztalos2010,ADMX:2018gho,ADMX:2019uok,ADMX:2018ogs,Crisosto:2019fcj,Lee:2020cfj,Jeong:2020cwz,CAPP:2020utb,Devlin:2021fpq,HAYSTAC:2018rwy,HAYSTAC:2020kwv,McAllister:2017lkb,Alesini:2019ajt,Alesini:2020vny,CAST:2021add,DePanfilis,Hagmann}, and helioscope searches for solar axions~\cite{CAST:2007jps,CAST:2017uph}. We stress that these bounds are intended for illustration and context only, and we have not yet recast them into the companion-axion parameter space (see later figures).} 
\label{fig:Couplings}
\end{center}
\end{figure} 

When additional sources of PQ symmetry breaking are present, the effective axion potential gains another term,
\begin{equation}\label{pot1}
V(a)=-2K\cos \left(N\frac{a}{f_a}+\theta\right)-2\kappa K\cos \left(N_g\frac{a}{f_a}+\theta_g\right)~,
\end{equation}
where $N$ and $N_g$ denote model-dependent anomaly coefficients, and $f_a$ is the scale of PQ symmetry breaking. Here $K=\frac{m_um_d}{(m_u+m_d)}m^2_{\pi}f^2_{\pi}$ measures the contribution coming from QCD instantons, whereas the parameter $\kappa$ quantifies the relative strength of the additional contribution, estimated to be $\kappa\sim 0.04-0.6$ for gravitational instantons~\cite{Chen:2021jcb}.

Since the additional CP-violating $\theta_g$ parameter is unrelated to $\theta$ and \emph{a priori} should be $\sim \mathcal{O}(1)$, the vacuum configuration of the single axion field is no longer able to cancel the undesired CP violating effects. Hence, axion models \cite{Weinberg:1977ma, Wilczek:1977pj, Kim:1979if, Shifman:1979if, Zhitnitsky:1980tq, Dine:1981rt} are not valid solutions to the strong CP problem unless the additional contribution is sufficiently small, $\kappa < 10^{-9}$~\cite{Chen:2021jcb}. 

Perhaps the simplest way to solve this new CP problem is to propose another PQ symmetry, implying the existence of a `companion' axion in the particle spectrum. Depending on their relative PQ scales, the solution presents the tantalizing possibility that \emph{two} QCD axions could be seen in future experiments. In this paper, we begin an investigation into the phenomenology of this model by examining the detectability of the two axions via their couplings to the photon.

Before proceeding, we point out that multi-axion models are not a radically new idea. In particular, the string axiverse scenario~\cite{Masso:1995tw, Masso:2002ip, Ringwald:2012hr, Ringwald:2012cu, Arvanitaki:2009fg, Svrcek:2006yi, Acharya:2010zx,Cicoli:2012sz, Jaeckel:2010ni,Stott:2017hvl} has inspired many recent studies of systems of coupled axions~\cite{Kitajima:2014xla,Marsh:2019bjr,Cyncynates:2021yjw,Reig:2021ipa,Chadha-Day:2021uyt}. Additional axions have also been introduced in the context of inflation~\cite{Kim:2004rp,Dimopoulos:2005ac}, to enhance~\cite{Agrawal:2017cmd} or suppress~\cite{Babu:1994id,Dror:2020zru,Dror:2021nyr} couplings to the Standard Model, to explain astrophysical anomalies~\cite{Higaki:2014qua}, or as relics of some other high-energy physics~\cite{Hu:2020cga}. The `companion axion' model that we investigate here is distinct in that it requires two coupled particles to solve the strong-CP problem. This will significantly restrain the parameter space compared to axiverse models.

\textbf{\textit{The companion axion model}}.---Proposed in~\cite{Chen:2021jcb}, the companion-axion model protects the PQ solution to the strong CP problem from colored Eguchi-Hanson (CEH) instantons~\cite{Boutaleb-Joutei:1979vmg} by extending the original $U(1)_{\rm PQ}$ symmetry to $U(1)_{\rm PQ}\times U(1)'_{\rm PQ}$. Once spontaneously broken, two pseudo-Goldstone axions appear, with the potential 
\begin{align}
V(a, a')=&-2K\cos \left(N\frac{a}{f_a}+N' \frac{a'}{f_a'}+\theta\right)\nonumber \\
&-2\kappa K\cos \left(N_g\frac{a}{f_a}+N'_g \frac{a'}{f_a'}+\theta_g\right)~. 
\label{pot2}
\end{align}
The lowest energy state is then realized for the axion field expectation values that cancel out both CP-violating terms. As long as $NN'_g\neq N'N_g$, the strong CP problem can be dynamically resolved \emph{\`{a} la} Peccei-Quinn.

The two axion states $a$ and $a'$ are mixed through the interactions in Eq.(\ref{pot2}). The mixing angle $\alpha$ between mass eigenstates $a_1$ and $a_2$ is, 
\begin{equation}
    \tan 2\alpha = \frac{2\epsilon(NN'+\kappa N_gN'_g)}{(N^2+\kappa N^2_g) - \epsilon^2 (N'^2+\kappa N'^2_g)} \,
\end{equation}
where for convenience we parameterize the ratio of the two PQ scales, $\epsilon \equiv f_a/f'_a$. The mass eigenstates are given by,
\begin{align}
    &m_{1}^2=\frac{\Delta m^2}{2}+\frac{K}{f^2_a}\bigg((N^2+\kappa N^2_g)+\epsilon^2(N^2+\kappa N^2_g)\bigg), \\
    &\Delta m^2 = \frac{2K}{f^2_a}\bigg[4(NN'+\kappa N_gN'_g)^2\epsilon^2\nonumber\\
    &~~~~~~~~~~~~~~~~~~~~+\Big((N^2+\kappa N^2_g)-\epsilon^2(N'^2+\kappa N'^2_g)\Big)^2\bigg]^{1/2}
\end{align} 
where $\Delta m^2=m_1^2-m_2^2>0$, making $a_1$ always the heaviest axion. Without loss of generality we can assume $\epsilon\leq 1$. 

The above expressions simplify significantly in two particular cases: a hierarchical regime, $f'_a\gg f_a$, and a strong-mixing regime, $f'_a\simeq f_a$.

\noindent {\it 1. Hierarchy}: $\epsilon\ll 1$. Assuming all anomaly coefficients are of the same order, and ignoring terms $\sim\mathcal{O}(\kappa^2)$, we obtain
\begin{align}
\label{massh1}
&m_1^2\approx \frac{2K\left(N^2+\kappa N_g^2\right)}{f_a^2}~, \\
\label{massh2}
&m_2^2\approx \frac{\kappa \left(NN_g'- N_g N'\right)^2}{N^4+\kappa N^2N_g^2}\epsilon^2m_1^2\sim \kappa \epsilon^2m_1^2 \, .
\end{align}
The heavier axion [Eq.(\ref{massh1})] has a mass the same order as the standard QCD axion, while the mass of the second axion [Eq.(\ref{massh2})] is determined by the relative size of the additional CEH instanton contribution ($\sim \kappa$) and has a further $\epsilon$ suppression due to its higher PQ scale. The mixing between the two axions in this regime is small because,
\begin{equation}
   \alpha \approx \frac{NN'+\kappa N_gN_g'}{N'^2+\kappa N_g'^2} \epsilon\sim \epsilon\ll 1 \, .
\label{angleh}  
\end{equation}

\noindent {\it 2. Mixing}: $\epsilon\simeq 1$. In this case we have instead,
\begin{align}
\label{massm1}
&m_1^2\approx \frac{2K}{f_a^2} \left[N^2+N'^2+\kappa\frac{N^2N_g^2+N'^2N_g'^2}{N^2+N'^2}\right]~, \\
\label{massm2}
&m_2^2\approx \frac{2K\kappa}{f_a^2}~\frac{N^2N_g'^2+N'^2N_g^2}{N^2+N'^2}\sim \kappa m_1^2~
\end{align}
Once again, the mass of the heavier axion [Eq.(\ref{massm1})] is of the same order of magnitude as the mass of the standard QCD axion, while the mass of the lighter axion [Eq.(\ref{massm2})] is entirely defined by the CEH instantons. Unlike the hierarchical regime, the axions in this case are strongly mixed,
\begin{equation}
    \tan2\alpha\approx -\frac{2\left(NN'+\kappa N_gN_g'\right)}{\left(N^2-N'^2\right)+\kappa\left(N_g^2-N_g'^2\right)}~.
\label{anglem} 
\end{equation}

\begin{figure*}[t]
\begin{center}
\includegraphics[trim = 0mm 0mm 0mm 0mm, clip, height=0.45\textwidth]{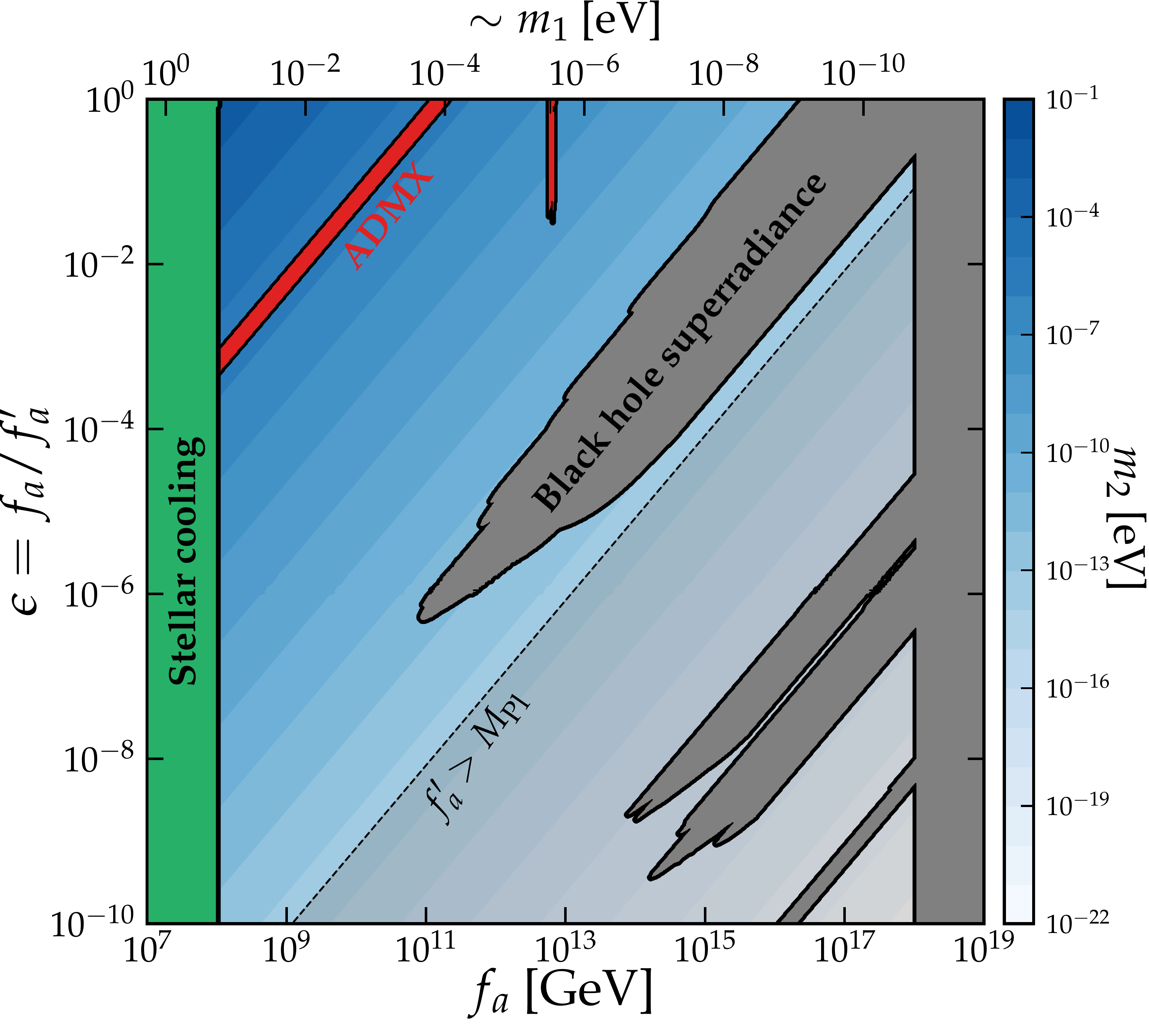}\quad\quad
\includegraphics[trim = 0mm 0mm 0mm 0mm, clip, width=0.445\textwidth]{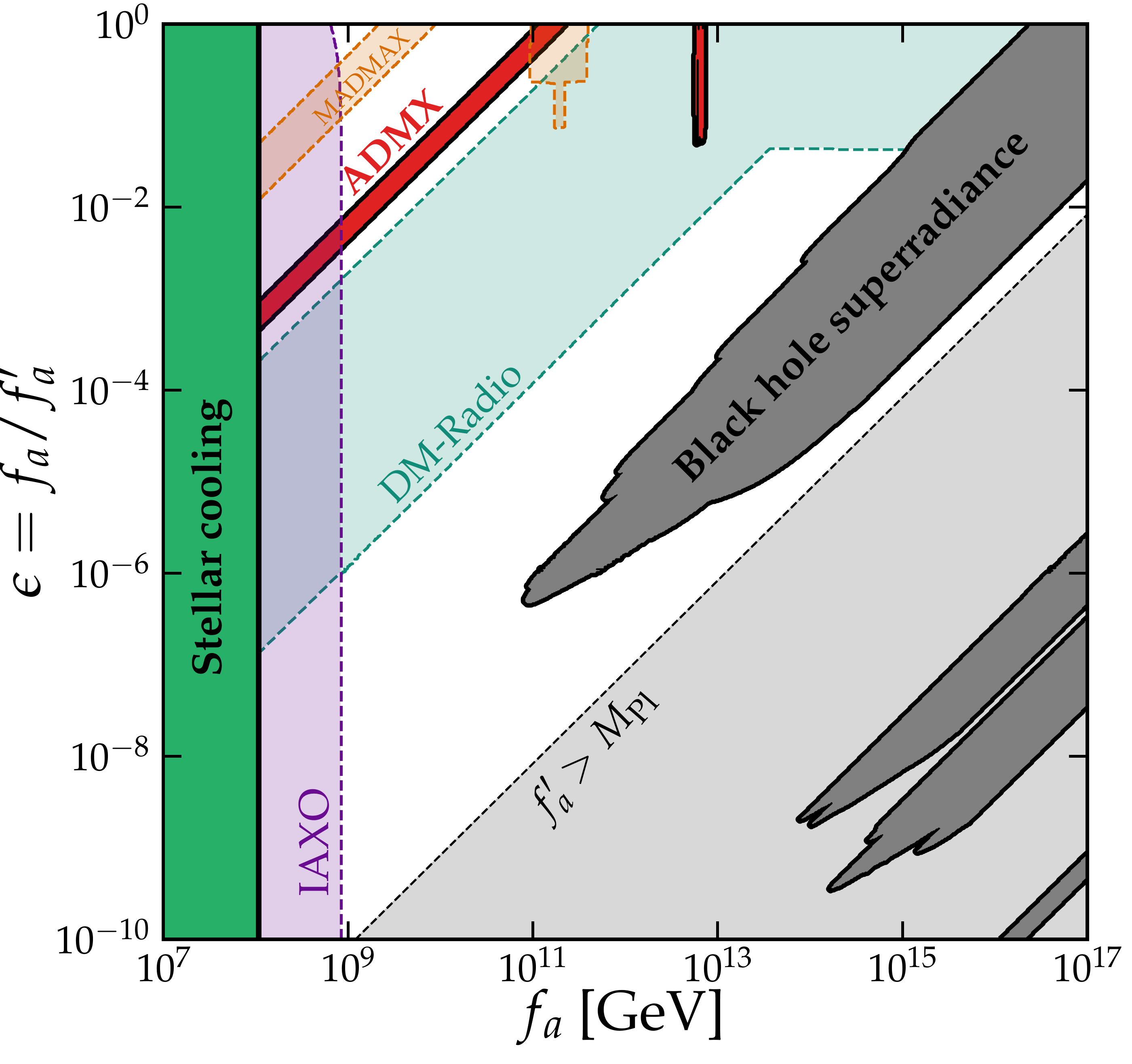}
\caption{{\bf Left:} Current bounds on the companion-axion model. The colorscale corresponds to the value of the lighter axion's mass, whereas the heavier axion's mass is shown (roughly) by the upper horizontal axis. We can rule out parts of this parameter space using stellar cooling arguments, ADMX, and black hole superradiance. {\bf Right:} As in the left-hand panel, but now showing projected constraints from future experiments: MADMAX~\cite{Beurthey:2020yuq}, IAXO~\cite{Armengaud:2019uso} and DMRadio/ABRACADABRA~\cite{DMRadio,Kahn:2016aff}.} 
\label{fig:AxionMasses}
\end{center}
\end{figure*}

\textbf{\textit{Axion-photon couplings}}.---The couplings of the two axions to Standard Model fields can be computed using the usual techniques~\cite{Srednicki:1985xd, GrillidiCortona:2015jxo, DiLuzio:2020wdo}. To minimize the number of free parameters, we assume that the UV completion of the two-axion model is similar to the KSVZ model, popular in the single-QCD-axion case~\cite{Kim:1979if, Shifman:1979if}. As such, the axions couple only to electrically neutral super-heavy quarks that carry charges under the extended $U(1)_{\rm PQ}\times U(1)'_{\rm PQ}$ Peccei-Quinn symmetry. We then take suitable linear combinations of anomalous Peccei-Quinn and light-quark chiral currents and extract couplings of axion states $a$ and $a'$ to the photon:     
\begin{align}
\mathcal{L}_{a\gamma}=\frac{1}{4}\left(a g_{a\gamma}+a' g'_{a\gamma}\right)F_{\mu\nu}\tilde F^{\mu\nu}~,
\label{Lax-ph} 
\end{align}
where, 
\begin{align}
&g_{a\gamma}=g'_{a\gamma}\frac{f'_a}{f_a}\frac{N}{N'}=-\frac{\alpha_{\rm em}N}{2\pi f_a}\zeta,~~~
\zeta =\frac{2}{3}\frac{4m_d+m_u}{m_u+m_d}~,
\label{ax-ph} 
\end{align}
where the factor $\zeta\approx 1.92$ is fixed to avoid the mixing of axions with QCD mesons \cite{Bardeen:1977bd}. The photon couplings to the axion mass eigenstates can be readily obtained via, 
\begin{align}\label{couplingeq}
    g_1&=\frac{\alpha_{\rm em}\zeta}{2\pi f_a}(N\cos\alpha-\epsilon N'\sin\alpha)~,\\
    g_2&=\frac{\alpha_{\rm em}\zeta}{2\pi f_a}(N\sin\alpha+\epsilon N'\cos\alpha) \, .
\end{align}

The couplings $g_{1,2}$ as a function of $m_{1,2}$ are shown in Fig.~\ref{fig:Couplings}. We choose representative values of the anomaly coefficients: $\{N,N^\prime,N_g,N^\prime_g\} = \{3,1/2,13/2,3/2\}$\footnote{In KSVZ-like models the $SU(3)_c$ representations of the two sets of heavy quarks have to be different. We normalize Peccei-Quinn charges to unity.}, and an instanton contribution ratio conservatively set to $\kappa = 0.04$~\cite{Chen:2021jcb}. The precise numbers are unimportant for our qualitative conclusions, but can lead to non-trivial quantitative differences to the couplings, including the possibility of cancellations.

In Fig.~\ref{fig:Couplings} we choose $\epsilon = f_a/f^\prime_a = 10^{-1}$ such that the two axions are in the mixing regime and both of their masses can be displayed on the same plot. We can see that $a_1$ lies along the standard KSVZ line, whereas $a_2$ is always lighter. To provide familiar context, we have overlaid existing haloscope and helioscope constraints on the \emph{single-axion}. The remainder of this work, however, will focus on re-deriving these constraints under the companion-axion model, including effects that are distinct from the single-axion case.

In our model we have two QCD axions and therefore two free parameters. Since we have fixed $f_a\leq f^\prime_a$, we can map existing and future constraints on our model by defining the parameter space $(f_a,\epsilon)$, shown in Fig.~\ref{fig:AxionMasses}. We use the color scale to encode the mass of the lighter state, $m_2$, where we see that constant values of $m_2$ roughly correspond to contours $\epsilon\propto f_a$. The mass of the heavier state $m_1$, in the hierarchical approximation [Eq.(\ref{massh1})], is shown by the upper horizontal axis. We now describe how we have derived each bound shown in this figure.

\textbf{\textit{Stellar cooling}}.---Stars can be powerful factories for axions with masses smaller than their internal temperatures $\mathcal{O}($keV)~\cite{Raffelt:2006cw}. The most stringent bound on the photon coupling comes from the numbers of horizontal branch stars in globular clusters, whose lifetimes are sensitive to Primakoff production of axions $\gamma+Z e \rightarrow Z e+a$. Subsequently, single-axion-photon couplings,
\begin{equation}\label{eq:boundg}
  \left|g_{a \gamma}\right|<6.6 \times 10^{-11} \, {\rm GeV}^{-1}\, ,
\end{equation}
are ruled out at 95\% C.L.~\cite{Ayala:2014pea}. This bound holds independently of the axion mass up to around $m_a\approx 30$~keV. We can convert this into a bound on the two-axion model by replacing $g_{a\gamma}$ in Eq. \eqref{eq:boundg} with the electromagnetically active coupling combination $\tilde{g}=\sqrt{g_1^2+g_2^2}$ (see below). As would be expected, in the hierarchical case ($g_2\ll g_1)$ the bound is the same as the single axion model because the lighter axion is not efficiently produced; whereas when $f_a \approx f'_a$, both axions are generated by the star and the bound is enhanced by a factor $\sqrt{N^2+N'^2}/N$.

\textbf{\textit{Helioscopes}}.---A helioscope~\cite{Sikivie:1983ip} consists of a long magnetic bore pointed directly at the Sun, with a system of X-ray optics and detectors placed at the opposite end to capture solar axions converting into photons. For very light masses, the axion and photon oscillate coherently along the length of the magnet---in CAST, for example, their vacuum-mode limit holds for $m_a\lesssim 0.02$~eV. Above this mass, the momentum mismatch between the massive axions and the massless photon generates oscillations in the conversion probability over length-scales shorter than the experiment, suppressing the observable signal. The mass reach can be extended to the QCD axion band by providing the photon with a variable plasma mass which permits resonant conversion whenever the photon mass matches the axion mass---in CAST this is done by filling the bore with helium~\cite{CAST:2013bqn,CAST:2015qbl}.

For helioscopes, unlike the stellar bounds, we must also consider the axion propagation to the detector. The two axions can be written in a basis where one particle state is an electromagnetically `active' sum of the two axions, and the other state is `hidden'~\cite{Chadha-Day:2021uyt}. The mixing angle of this system is given by,
\begin{align}\label{activemix}
    \cos\theta = \frac{g_1}{\sqrt{g_1^2+g_2^2}}~,
\end{align}
and the resulting off-diagonal term $M_{12}$ in the axion mass matrix is,
\begin{equation}\label{offdiag}
    M_{12} = -\Delta m^2\frac{\cos^2\alpha}{1+\epsilon^2}\Big(\tan\alpha+\epsilon\big(1-\tan^2\alpha\big)-\epsilon^2\tan\alpha\Big)  
\end{equation} 
While the Sun only emits the electromagnetically active state, the axions propagate in the mass-basis and so oscillate as they travel to Earth (see e.g.~\cite{Chadha-Day:2021uyt}). Ultimately this will reduce the observable portion of the axion flux. The survival probability of the active axion with energy $\omega$, after travelling a distance $L$ is,
\begin{align}\label{eq:survivalprob}
    P &= 1-\sin^2 2\theta \sin^2 \left( \frac{ \Delta m^2 L}{4\omega}\right)~,
\end{align}
where the first sine can be written in terms of our model parameters using Eqs.(\ref{couplingeq}) and (\ref{activemix}),
\begin{align}
    \sin^2 2\theta &= \frac{4\cos^4 \alpha \left(\tan \alpha + \epsilon \left(1-\tan^2 \alpha \right)-\epsilon^2\tan \alpha\right)^2}{\left(1+\epsilon^2\right)^2}\nonumber\\
    &\approx \begin{cases}
        4\epsilon^2, & \epsilon\ll 1 \\
        \cos^2 2\alpha, & \epsilon\approx 1
        \end{cases}~.
\end{align} 
Taking $L = 1$~AU, and $\omega \sim $~keV, we can see that when $\Delta m^2 \gtrsim 10^{-12}$~eV$^2$, or equivalently, $f_a \lesssim 10^{13}$~GeV, the axion-axion oscillation length is shorter than the Earth-Sun distance. As a result, the conversion probability will oscillate rapidly as a function of $\omega$, and the second sine in Eq.(\ref{eq:survivalprob}) averages to $1/2$. Since even next-generation helioscopes will only be sensitive to $f_a\lesssim10^9$~GeV, we can assume we are always within this averaged regime.


Once the axions arrive at Earth they enter the strong transverse magnetic field of the helioscope, $B$, where the active axion state can now also oscillate into photons with $\tilde{g}B \omega$. While the three-particle oscillation problem is hard to solve analytically, we find that the axion-photon coupling is small compared to the coupling between the active and hidden axion. In the hierarchical case, $\tilde{g}=g\sqrt{1+\epsilon^2)}\approx g$ and we can use Eq.(\ref{offdiag}) to find the ratio of these couplings,
\begin{equation}\label{oscfact}
    \frac{\tilde{g}B\omega}{\Delta m^2\epsilon}=\frac{\alpha_{\rm em}\zeta B\omega}{4\pi K}f'_a\lesssim 10^{-2},
\end{equation} 
where the upper limit is set by taking $f'_a<M_{\rm pl}$.
A similar relation holds in the strong-mixing regime. Since we are well within the regime where the two axions are mixed, we can therefore approximate the detected photon flux just as we would in the single-axion case, but reduced by the fraction of the population in the `active' state. Any additional effects coming from the three-particle oscillation inside the experiment will be suppressed by the factor Eq.(\ref{oscfact}). Since the number of photons observed in a helioscope scales $\propto g_{a\gamma}^{4}$, we can recast existing and projected bounds by multiplying the minimum detectable photon coupling by the factor $P^{-1/4}$. 

\begin{figure}[t]\label{smokingfig}
\begin{center}
\includegraphics[trim = 0mm 0mm 0mm 0mm, clip, width=0.47\textwidth]{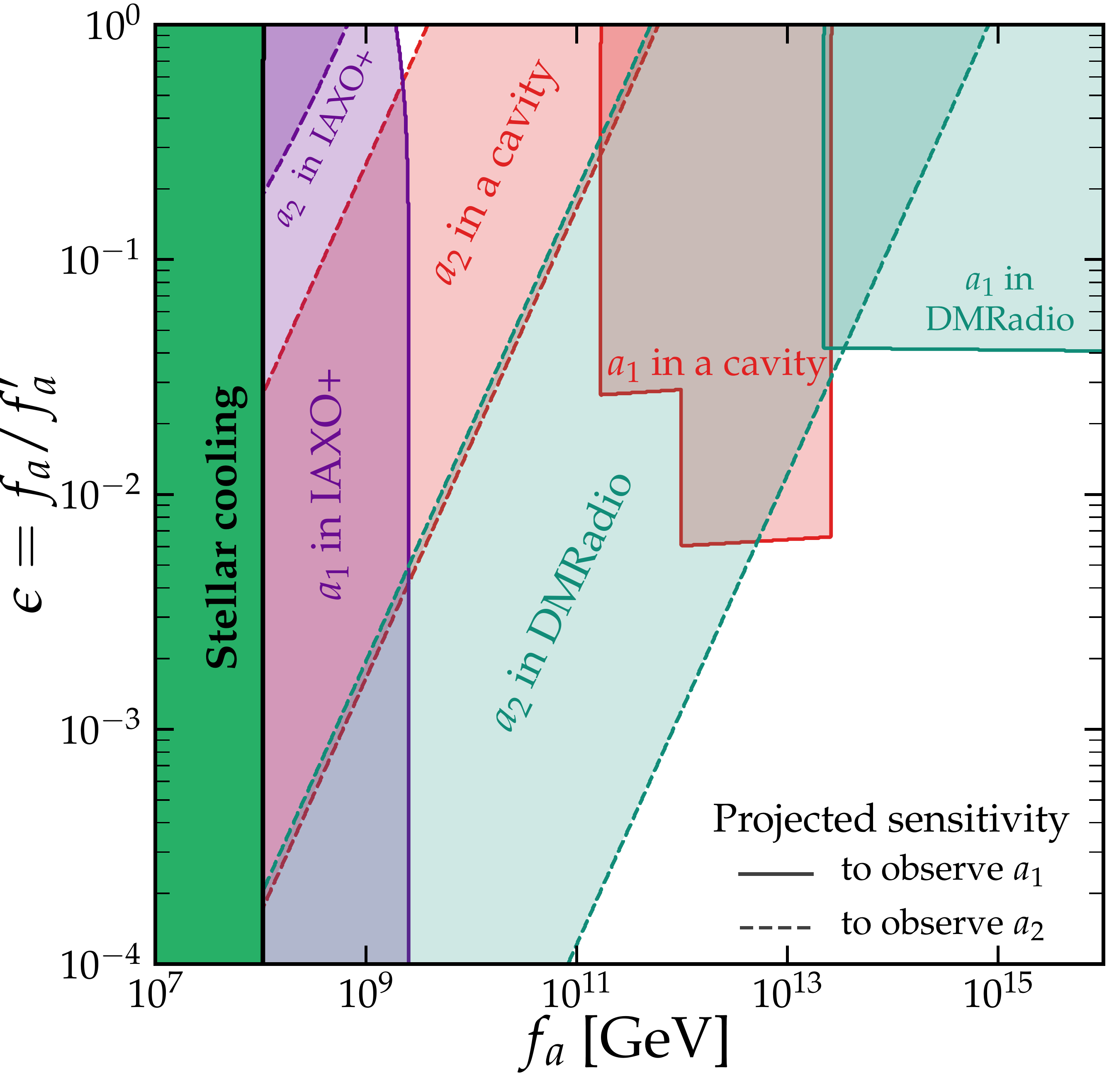}
\caption{Projected sensitivity for loosely-defined future experiments: an upgraded version of IAXO~\cite{Armengaud:2018cuy}, resonant cavity experiments covering the 1--100~$\upmu$eV region (based on projections from~\cite{Stern:2016bbw}), and DMRadio-like experiments covering the neV--$\upmu$eV region (based on~\cite{DMRadio}). In the parts where the $a_1$ and $a_2$ regions overlap, the signal from \emph{both} axions will be seen---a characteristic of the companion axion model. For some small regions the signal of the two axions could even be seen in the same experiment.}
\label{fig:Projections}
\end{center}
\end{figure}

Just as with the single-axion, the CAST bounds are less sensitive than the stellar cooling bounds, so they do not appear on our plot. Instead, we show the projected bounds for the future helioscope IAXO~\cite{Armengaud:2019uso}. The result can be observed in Fig.~\ref{fig:Projections}, where we see that IAXO is expected to improve upon the stellar bound by just under an order of magnitude in $f_a$, and with a limit that is mostly insensitive to $\epsilon$---apart from in the strong-mixing regime where the axion-axion oscillations slightly impact the detectable flux. For this particular model configuration both axions could be seen during IAXO's resonant buffer gas phase, however it may still be possible to measure the mass even in the vacuum phase if $a_2$ turned out to be lighter~\cite{Dafni:2018tvj}.

\textbf{\textit{Haloscopes}}.---Axion haloscopes~\cite{Sikivie:1983ip} aim to detect axions constituting the DM halo of the Milky Way. In principle, rescaling past bounds set by experiments such as ADMX should be simple once we know the ratio of the local DM that is comprised of each axion. In the single axion case one can assume $\Omega_a = \Omega_{\rm dm}$ and remain agnostic towards how those axions were created. However, with two axions we are forced into making additional assumptions about their respective production mechanisms.

There are already complications involved in making a clear prediction for the cosmological abundance of axions when there is only one scale to deal with: including whether that scale is higher or lower than the scale of inflation, as well as the effects of topological defects~\cite{Kawasaki:2014sqa,Fleury:2015aca,Klaer:2017ond,Buschmann:2019icd,Vaquero:2018tib,Gorghetto:2018myk,Gorghetto:2020qws,Buschmann:2021sdq}. In the companion axion scenario---where we have two scales---the situation is naturally more complicated. A more complete study of companion axion DM production in the early universe is therefore deserving of a full study~\cite{newpaper} (see also the recent~\cite{Cyncynates:2021yjw}). However, to make progress we can adopt a crude estimate for the proportions of the DM made up of each axion. Assuming production by the misalignment mechanism alone (neglecting axions generated by topological defects), an order of magnitude estimate of the density ratio is,
\begin{align}
    &\frac{\Omega_{a_2}}{\Omega_{a_1}}\sim\frac{\theta^2_2}{\theta_1^2}\kappa^{0.41}\epsilon^{-1.19},
\end{align} 
valid for temperature-dependent axion masses, $m^2(T)\sim T^{-n}$, with $n=6.68$ \cite{Wantz:2009it}. Unless $\epsilon\approx1$ we expect the relic abundance to be dominated by the lighter axion. For our estimates we assume $\theta_1=\theta_2$, but if one or both of the PQ symmetries are broken before inflation these angles could be tuned to other values by anthropic arguments. Therefore, since our haloscope bounds are contingent on one particular cosmological scenario, they should be regarded as more model-dependent than single-axion bounds.

The companion-axion model offers an intriguing prospect for resonance-based haloscopes which must scan slowly across a mass range to search for a signal. Depending on $\Omega_{a_{1,2}}$, there could be a sizeable signal to be discovered at two distinct frequencies. But even if only one axion falls within reach of some experiment, a combined constraint on the companion-axion model can still be made. Since the model poses that both axions must exist, a constraint on a particular value of $f_a$ immediately implies that some $f^\prime_a = f_a/\epsilon$ is excluded as well. As such, we expect that a small section of some single-axion model that is ruled out by a haloscope to translate into two bands of ruled-out models in the $(f_a,\epsilon)$ space, as in Fig.~\ref{fig:AxionMasses}. ADMX currently excludes KSVZ axions around $m_a\sim3\times 10^{-6}$~eV. When $\epsilon\gtrsim10^{-2}$, ADMX excludes this same window because it would have seen the heavier axion there. However ADMX \emph{also} excludes a diagonal band of smaller $f_a$ in the hierarchical regime, where the lighter axion would have been observed instead.

\textit{\textbf{Black hole superradiance}}.---If light bosonic fields exist then they can form an exponentially growing bound state around a spinning BH. For fields with Compton wavelengths around the size of the BH ergoregion, the mechanism known as superradiance can act to extract the entirety of the BH's spin~\cite{Dolan:2007mj, Arvanitaki:2010sy, Pani:2012vp, Brito:2015oca, Arvanitaki:2014wva, Arvanitaki:2016qwi, Herdeiro:2016tmi, Cardoso:2018tly, Stott:2018opm}. This means that the measurement of a BH spin can serve to rule out both axions if they have masses within the appropriate range.

The literature on this subject is developing, and there is not resounding agreement between the bounds derived by various groups~\cite{Baryakhtar:2020gao,Baryakhtar:2017ngi,Stott:2020gjj,Mehta:2020kwu,Mehta:2021pwf} who differ in both their theoretical and statistical treatments. We adopt the constraints derived using techniques described in Refs.~\cite{Stott:2020gjj,Mehta:2020kwu,Mehta:2021pwf} which cover both stellar and supermassive BHs. We assume that superradiance rules out both axion mass eigenstates but only when $f_a$ or $f^\prime_a$ are not below the scale where the bounds relax due to self-interactions. The four bands in Fig.~\ref{fig:AxionMasses} correspond to various stellar-mass and supermassive BHs with measured spins. We note however that different, more conservative bounds have also been derived~\cite{Baryakhtar:2020gao} using a more involved treatment of the field's self-interactions, and neglecting supermassive BH spins which have been measured only at low significance. We should note also that the formation of the two axions' bound states may proceed differently if there is mixing between the two eigenstates with different masses. Hence the bounds for $\epsilon$ close to 1 may not be accurate---although a detailed calculation of superradiance in this model is not our focus.

\textbf{\textit{Discovering the companion axion model}}.---While for much of our parameter space one of the axions is rather light and weakly coupled, over the next few decades, plans are in place for experiments to scan almost the entirety of the QCD axion model band~\cite{Irastorza:2018dyq}. Therefore, we finish by estimating how much of our companion-axion parameter space would leave a \emph{unique} signal in experiments. 

In Fig.~\ref{fig:Projections} we show the sensitivities of IAXO+~\cite{Armengaud:2019uso,Armengaud:2014gea,IAXO:2020wwp}, as well as resonant-cavity~\cite{McAllister:2017lkb,Stern:2016bbw,Melcon:2018dba,AlvarezMelcon:2020vee,Alesini:2017ifp,Jeong:2017hqs} and LC-circuit-based haloscopes~\cite{Kahn:2016aff,DMRadio,Devlin:2021fpq,Ouellet:2018beu,Crisosto:2019fcj,Gramolin:2020ict,Salemi:2021gck}. We opt here for the most ambitious projections that have been made---ones which essentially cover all of the QCD band above $m_a\sim$~neV. This space could also be covered by non-cavity haloscopes~\cite{TheMADMAXWorkingGroup:2016hpc,Schutte-Engel:2021bqm,BRASS,Lawson:2019brd,Baryakhtar:2018doz}, which we neglect to reduce clutter.

Helioscopes cannot probe above $f_a\sim10^9$~GeV, meaning that some assumption about DM will be needed to explore further. For small $\epsilon$ and large $f_a$ there is a challenging region where no proposed experiments are sensitive. This is because the axion that dominates the DM abundance here is too weakly-coupled to detect in any proposed haloscope, and the low DM density in the heavier axion makes its signal too small to detect. It seems unlikely that experiments exploiting alternative couplings would be able to reach this regime either, but it may be possible to explore it via gravitational signatures~\cite{Hook:2017psm,Zhang:2021mks}.
 
\textbf{\textit{Conclusions}}.---Colored gravitational instantons jeopardize the single-axion solution to the strong-CP problem. A potential remedy, suggested by Ref.~\cite{Chen:2021jcb}, is to include a second ``companion'' axion that acts to remove the additional unwanted CP-violation. One axion is similar to the usual QCD axion that is being actively sought in experiments, whereas its companion would be present at some lighter mass. Both axions exist around the conventional QCD band, so the model would not demand any alterations to ongoing axion search campaigns. In fact, we predict that the signal of two axions may well appear, either in two different experiments, or perhaps even in the same experiment. One of the most remarkable messages that can be taken from this result is that even if an experiment does identify the signal of an axion, the remaining experiments operating at different frequencies should continue to search.

We have only delved into the implications for the photon coupling, which is just one dimension of the axion's rich phenomenology. We anticipate many more interesting signals unique to the companion-axion model: for example via couplings to fermions~\cite{Mitridate:2020kly,Chigusa:2020gfs,Ikeda:2021mlv,QUAX:2020adt,Crescini:2018qrz,Aybas:2021nvn,Garcon:2019inh,JacksonKimball:2017elr,Abel:2017rtm,OHare:2020wah,Arvanitaki:2014dfa,JacksonKimball:2017elr}, astrophysical signatures~\cite{Hook:2017psm,Hook:2018iia,Dessert:2021bkv}, or cosmological behavior~\cite{newpaper}. All of these may assist in either ruling out the remaining parameter space, or lead to an eventual discovery. 

The figures from this article can be reproduced using the code available at \url{https://github.com/cajohare/CompAxion}, whereas the data for all the limits shown here is compiled at Ref~\cite{AxionLimits}.

\textbf{\textit{Acknowledgements}}.---CAJO thanks Viraf Mehta and David Marsh for making available their superradiance constraints. The work of AK was partially supported by the Australian Research Council through the Discovery Project grant  DP210101636 and by the Shota Rustaveli National Science Foundation of Georgia (SRNSFG) through the grant DI-18-335.
\bibliography{axions.bib}
\bibliographystyle{bibi}
\end{document}